\documentclass{llncs}

\usepackage[vlined,boxed]{algorithm2e}

\usepackage{amsmath}
\usepackage{amssymb}
\usepackage{amstext}
\usepackage{tikz}
\usetikzlibrary{shapes,arrows}
\usepackage{alltt}
\usepackage{tikz}
\usetikzlibrary{shapes,arrows}
\usepackage{enumitem}
\usepackage{subfigure}

\begin{document}

\tikzstyle{decision} = [diamond, draw,
    text width=3em, text badly centered, inner sep=0pt, aspect=2, minimum width=4.1em, minimum height=2.3em]
\tikzstyle{block} = [rectangle, draw,
    text width=3em, text centered, minimum height=2em, minimum width=4.1em]
 \tikzstyle{nobblock} = [rectangle, draw,
    text width=3em, text centered, minimum height=2em]
\tikzstyle{line} = [draw, color=black, -latex]
\tikzstyle{block2} = [rectangle, draw,
    text width=3em, text centered, rounded corners, minimum height=2em]
\tikzstyle{linewithout} = [draw, color=black]
\tikzstyle{cloud} = [draw, ellipse]

\title{A Simple and Scalable Static Analysis for Bound Analysis and Amortized Complexity Analysis}

\author{Moritz Sinn \and Florian Zuleger \and Helmut Veith
\thanks{Supported by the Austrian National Research
    Network S11403-N23 (RiSE) of the Austrian Science Fund (FWF) and
    by the Vienna Science and Technology Fund (WWTF) through
    grants PROSEED and ICT12-059.}
}

\institute{
TU Vienna
}

\maketitle

\vspace*{-4mm}
\begin{abstract}
We present the first scalable \emph{bound analysis} that achieves \emph{amortized complexity analysis}.
In contrast to earlier work, our bound analysis is not based on general purpose reasoners such as abstract interpreters, software model checkers or computer algebra tools.
Rather, we derive bounds directly from abstract program models, which we obtain from programs by comparatively simple invariant generation and symbolic execution techniques.
As a result, we obtain an analysis that is more predictable and more scalable than earlier approaches.
We demonstrate by a thorough experimental evaluation that our analysis is fast and at the same time able to compute bounds for challenging loops in a large real-world benchmark.
Technically, our approach is based on lossy vector addition systems (VASS).
Our bound analysis first computes a lexicographic ranking function that proves the termination of a VASS, and then derives a bound from this ranking function.
Our methodology achieves amortized analysis based on a new insight how lexicographic ranking functions can be used for bound analysis.
\end{abstract}

\newcounter{counter}

\newcommand{\true}{\ensuremath{\mathit{true}}}
\newcommand{\false}{\ensuremath{\mathit{false}}}
\newcommand{\dom}{\ensuremath{\mathit{Def}}}
\newcommand{\range}{\ensuremath{\mathit{Range}}}
\newcommand{\Id}{\ensuremath{\mathit{Id}}}

\newcommand{\prog}{\ensuremath{P}}
\newcommand{\vars}{\ensuremath{\mathit{Var}}}
\newcommand{\varsP}{\ensuremath{\vars^\prime}}
\newcommand{\loc}{\ensuremath{l}}
\newcommand{\locs}{\ensuremath{L}}
\newcommand{\edges}{\ensuremath{E}}
\newcommand{\trn}{\ensuremath{\rho}}
\newcommand{\trnAlt}{\ensuremath{\tau}}
\newcommand{\trns}{\ensuremath{\Gamma}}
\newcommand{\state}{\ensuremath{s}}
\newcommand{\states}{\ensuremath{\Sigma}}
\newcommand{\paath}{\ensuremath{\pi}}
\newcommand{\paathAlt}{\ensuremath{\mu}}
\newcommand{\paathL}{\ensuremath{\nu}}
\newcommand{\paths}{\ensuremath{\mathtt{paths}}}
\newcommand{\rel}{\ensuremath{\mathtt{rel}}}
\newcommand{\relS}{\ensuremath{\rel_S}}
\newcommand{\relW}{\ensuremath{\rel_W}}
\newcommand{\looop}{\ensuremath{\mathit{loop}}}
\newcommand{\header}{\ensuremath{\mathit{header}}}
\newcommand{\val}{\ensuremath{\sigma}}
\newcommand{\reach}{\ensuremath{\mathtt{mpi}}}
\newcommand{\update}{\ensuremath{d}}
\newcommand{\increment}{\ensuremath{c}}

\newcommand{\elimV}{\ensuremath{\mathit{Elim}}}
\newcommand{\elim}{\ensuremath{\triangleright}}
\newcommand{\elimL}{\ensuremath{\elim_\mathit{local}}}
\newcommand{\elimG}{\ensuremath{\elim_\mathit{global}}}
\newcommand{\transinv}{\ensuremath{\mathit{TransInv}}}
\newcommand{\dcs}{\ensuremath{\mathit{DC}}}
\newcommand{\dcsE}{\ensuremath{\dcs_\mathit{eager}}}
\newcommand{\dcsL}{\ensuremath{\dcs_\mathit{lazy}}}
\newcommand{\dcsF}{\ensuremath{\dcs_\mathit{frame}}}
\newcommand{\rfCands}{\ensuremath{\mathit{LocRfCnd}}}
\newcommand{\rfCandsE}{\ensuremath{\rfCands_\mathit{eager}}}
\newcommand{\rfCandsL}{\ensuremath{\rfCands_\mathit{lazy}}}
\newcommand{\rfs}{\ensuremath{\mathit{LocRf}}}
\newcommand{\summary}{\ensuremath{\mathcal{S}}}
\newcommand{\subst}{\ensuremath{\sigma}}
\newcommand{\substCond}{\ensuremath{\sigma}}
\newcommand{\atrns}{\ensuremath{\mathcal{T}}}
\newcommand{\atrnsA}{\ensuremath{\mathcal{S}}}
\newcommand{\diff}{\ensuremath{\delta}}
\newcommand{\resets}{\ensuremath{\mathit{Reset}}}
\newcommand{\Reach}{\ensuremath{\mathit{Reach}}}
\newcommand{\candidates}{\ensuremath{N}}

\newcommand{\SCCs}{\ensuremath{\mathit{SCCs}}}
\newcommand{\SCC}{\ensuremath{\mathit{SCC}}}
\newcommand{\bound}{\ensuremath{\mathit{b}}}
\newcommand{\lex}{\ensuremath{\mathit{l}}}
\newcommand{\rf}{\ensuremath{\mathit{r}}}
\newcommand{\rfA}{\ensuremath{\rf_1}}
\newcommand{\rfB}{\ensuremath{\rf_2}}
\newcommand{\RankFs}{\ensuremath{\mathit{R}}}
\newcommand{\mode}{\ensuremath{\mathit{mode}}}
\newcommand{\expr}{\ensuremath{\mathit{expr}}}
\newcommand{\dec}{\ensuremath{\mathit{dec}}}
\newcommand{\inc}{\ensuremath{\mathit{inc}}}
\newcommand{\CFA}{CA}

\newcommand{\loopus}{{\tt Loopus }}
\newcommand{\koat}{{\tt KoAT }}
\newcommand{\pubs}{{\tt PUBS }}
\newcommand{\rank}{{\tt Rank }}
\newcommand{\terminator}{{\tt T2 }}

\section{Introduction}
\label{sec:introduction}
Automatic methods for computing bounds on the resource consumption of programs are an active area of research~\cite{conf/popl/HofmannJ03,conf/popl/GulwaniMC09,conf/sas/AliasDFG10,conf/pldi/GulwaniZ10,conf/sas/ZulegerGSV11,conf/sas/Alonso-BlasG12,journals/tcs/AlbertAGPZ12,journals/toplas/HoffmannAH12,AlbertGM12}.
We present the first scalable \emph{bound analysis} for imperative programs that achieves \emph{amortized complexity analysis}.
Our techniques can be applied for deriving upper bounds on how often loops can be iterated as well as on how often a single or several control locations can be visited in terms of the program input.

The majority of earlier work on bound analysis has focused on mathematically intriguing frameworks for bound analysis.
These analyses commonly employ general purpose reasoners such as abstract interpreters, software model checkers or computer algebra tools and therefore rely on elaborate heuristics to work in practice.
In this paper we take an orthogonal approach that complements previous research.
We propose a bound analysis based on a simple abstract program model, namely \emph{lossy vector addition systems with states}.
We present a static analysis with four well-defined analysis phases that are executed one after each other:
program abstraction, control-flow abstraction, generation of a lexicographic ranking function and bound computation.

A main contribution of this paper is a thorough experimental evaluation.
We compare our approach against recent bounds analysis tools~\cite{conf/sas/AliasDFG10,journals/tcs/AlbertAGPZ12,AlbertGM12,conf/tacas/BrockschmidtEFFG14}, and show that our approach is faster and at the same time achieves better results.
Additionally, we demonstrate the scalability of our approach by a comparison against our earlier tool~\cite{conf/sas/ZulegerGSV11}, which to the best of our knowledge represents the only tool evaluated on a large publicly available benchmark of C programs.
We show that our new approach achieves better results while increasing the performance by an order of magnitude.
Moreover, we discuss on
this benchmark how our tool achieves amortized complexity analysis in real-world code.

Our technical key contribution is a new insight how lexicographic ranking functions can be used for bound analysis.
Earlier approaches such as~\cite{conf/sas/AliasDFG10} simply count the number of elements in the image of the lexicographic ranking function in order to determine an upper bound on the possible program steps.
The same idea implicitly underlies the bound analyses~\cite{conf/cav/GulavaniG08,conf/popl/GulwaniMC09,conf/pldi/GulwaniJK09,conf/pldi/GulwaniZ10,conf/sas/ZulegerGSV11,AlbertGM12,conf/tacas/BrockschmidtEFFG14}.
However, this reasoning misses arithmetic dependencies between the components of the lexicographic ranking function (see Section~\ref{sec:overview}).
In contrast, our analysis calculates how much a lexicographic ranking function component is increased when another component is decreased.
This enables amortized analysis.

\paragraph{Related Work.}
An interesting line of research studies the \emph{amortized analysis} of first-order functional programs (e.g.~\cite{conf/popl/HofmannJ03,journals/toplas/HoffmannAH12}) formulated as type rules over a template potential function with unknown coefficients; these coefficients are then found by linear programming.
It is not clear how to transfer this approach to an imperative setting.
Promising first steps for the amortized analysis of imperative programs are reported in~\cite{conf/sas/Alonso-BlasG12}.
Quantifier elimination is applied for simplifying a constraint system over template cost functions.
Since quantifier elimination is expensive, the technique does not yet scale to larger programs.

\emph{Lexicographic ranking functions} in automated termination analysis have been pioneered by Bradley et al. (see~\cite{conf/cav/BradleyMS05} and follow-up papers) who employ an elaborate constraint solving technique.
A recent paper experimentally compares termination analysis by lexicographic ranking and transition invariants~\cite{conf/tacas/CookSZ13} implemented on top of a software model checker.
\cite{conf/sas/AliasDFG10} iteratively constructs a lexicographic ranking function by solving linear constraint systems.
\cite{conf/cav/BrockschmidtCF13} is a hybrid of the approaches~\cite{conf/tacas/CookSZ13} and  \cite{conf/sas/AliasDFG10}.
\cite{conf/cav/BradleyMS05}, \cite{conf/tacas/CookSZ13} and \cite{conf/cav/BrockschmidtCF13} compute a lexicographic ranking function for a \emph{single} control location (i.e., one loop header) at a time, while the application of bound analysis requires to find a common lexicographic ranking function for \emph{all} control locations.
\cite{conf/sas/AliasDFG10} computes such a ranking function, but is limited to fairly small programs.
Our approach complements the cited approaches as it represents a simple and scalable construction of a lexicographic ranking function for all control locations.

\emph{Bound Analysis.}
The COSTA project (e.g.~\cite{journals/tcs/AlbertAGPZ12,AlbertGM12})
studies the extraction of cost recurrence relations from Java bytecode programs and proposes new methods for solving them with the help of computer algebra systems.
\cite{conf/cav/GulavaniG08} proposes to extend the polyhedra abstract domain with max- and non-linear expressions.
\cite{conf/popl/GulwaniMC09} introduces multiple counters and exploits their dependencies such that upper bounds have to be computed only for restricted program parts.
\cite{conf/pldi/GulwaniJK09} proposes an abstract interpretation-guided program transformation that separates the different loop phases such that bounds can be computed for each phase in isolation.
\cite{conf/pldi/GulwaniZ10} employs proof-rules for bound computation combined with disjunctive abstract domains for summarizing inner loops.
\cite{conf/sas/ZulegerGSV11} proposes a bound analysis based on the size-change abstract domain.
\cite{backward_symb_exec,conf/tacas/BrockschmidtEFFG14} discuss how to alternate between bound analysis and invariant analysis for the mutual benefit of the computed bounds and invariants.

\section{Motivation and Overview}
\label{sec:overview}

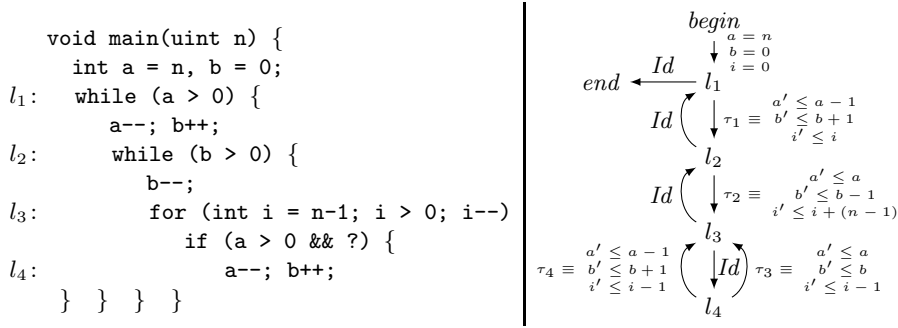
\begin{figure*}[t!]

\begin{tabular}{l|c}
\begin{minipage}[c]{6.8cm}
%
%
\small
\begin{alltt}

   void main(uint n) \{
     int a = n, b = 0;
\(l\sb{1}\!:\)   while (a > 0) \{
        a--; b++;
\(l\sb{2}\!:\)      while (b > 0) \{
           b--;
\(l\sb{3}\!:\)         for (int i = n-1; i > 0; i--)
              if (a > 0 && ?) \{
\(l\sb{4}\!:\)               a--; b++;
    \}  \}  \}  \}
\end{alltt}

\end{minipage}

&

\begin{minipage}[c]{5.4cm}
\begin{tikzpicture}[scale=0.4, node distance = 2cm, auto]

\node (t0)  {$\mathit{begin}$};
\node (t1) [below of=t0, node distance = 0.8cm] {$\loc_1$};
\node (t2) [below of=t1, node distance = 1cm]  {$\loc_2$};
\node (t3) [below of=t2, node distance = 1cm] {$\loc_3$};
\node (t4) [below of=t3, node distance = 1cm]  {$\loc_4$};
\node (t9) [left of=t1, node distance = 1.5cm]  {$\mathit{end}$};

\path
(t0) edge [line] node [right,font=\tiny ] {
 \begin{tabular}{c}
 $a = n$\\
 $b = 0$\\
 $i = 0$
 \end{tabular}
}(t1)
(t1) edge [line] node [right,font=\tiny ] {
$\trnAlt_1 \equiv$ \begin{tabular}{c}
 $a' \le a-1$\\
 $b' \le b+1$\\
 $i' \le i$
 \end{tabular}
}(t2)
(t2) edge [line] node [right,font=\tiny ] {
$\trnAlt_2 \equiv$ \begin{tabular}{c}
 $a' \le a$\\
 $b' \le b-1$\\
 $i' \le i+(n-1)$
 \end{tabular}
}(t3)
(t3) edge [line] node [right, xshift=-0.08cm, yshift=0.04cm] {$\Id$} (t4)
(t4) edge [line, out=30,in=-30] node [right,font=\tiny ] {
$\trnAlt_3 \equiv$ \begin{tabular}{c}
 $a' \le a$\\
 $b' \le b$\\
 $i' \le i-1$
 \end{tabular}
}(t3)
(t4) edge [line, out=150,in=-150] node [left,font=\tiny ] {
$\trnAlt_4 \equiv$ \begin{tabular}{c}
 $a' \le a-1$\\
 $b' \le b+1$\\
 $i' \le i-1$
 \end{tabular}
}(t3)
(t3) edge [line, out=150,in=-150] node [left] {$\Id$} (t2)
(t2) edge [line, out=150,in=-150] node [left] {$\Id$} (t1)
(t1) edge [line] node [above] {$\Id$} (t9);
\end{tikzpicture}
\end{minipage}

%
%
%

\end{tabular}
\caption{
Our running example, '?' denotes non-determinism (arising from a condition not modeled in the analysis). On the right we state the lossy VASS obtained by abstraction, $\Id$ denotes $a' \le a$, $b' \le b$, $i' \le i$.
}
\vspace*{-4mm}
\label{fig-ex1}
\end{figure*}

The example presented in Figure~\ref{fig-ex1} (encountered during our experiments) is challenging for an automated bound analysis:
(C1) There are loops whose loop counter is modified by an inner loop: the innermost loop modifies the counter variables $a$ and $b$ of the two outer loops.
Thus, the inner loop \emph{cannot be ignored} (i.e., cannot be sliced away) during the analysis of the two outer loops.
(C2) The middle loop with loop counter $b$ requires a \emph{path-sensitive} analysis to establish the linear loop bound $n$:
it is not enough to consider how often the innermost loop can be executed (at most $n^2$ times) but rather how often the if-branch of the innermost loop (on which $b$ is actually incremented) can be executed (at most $n$ times).
(C3) Current bound analysis techniques cannot model \emph{increments} and instead approximate increments by \emph{resets}, e.g., approximate the increment of $b$ by an assignment to a value between 0 and $n$ (using the fact that $n$ is an upper bound of $b$)!
Because of this overapproximation no bound analysis from the literature is able to compute the linear loop bound $n$ for the middle loop.
We now illustrate the main steps of our analysis:

\underline{\emph{1. Program Abstraction:}}
First, our analysis abstracts the program to the VASS depicted in Figure~\ref{fig-ex1}.
We introduce VASSs in Section~\ref{sec:program-model}.
In this paper we are using \emph{parameterized} VASSs, where we allow increments that are symbolic but constant throughout the program (such as $n-1$).
We extract lossy VASSs from C programs using simple invariant generation and symbolic execution techniques (described in Section~\ref{sec:abstraction}).

\underline{\emph{2. Control Flow Abstraction:}}
We propose a new abstraction for bound analysis, which we call \emph{control flow abstraction} (\CFA) (described in Section~\ref{sec:transition-system}).
\CFA\ abstracts the VASS from Figure~\ref{fig-ex1} into a transition system with four transitions:
$\trn_1 \equiv a' \le a-1 \wedge b' \le b+1 \wedge i' \le i, \quad
\trn_2 \equiv a' \le a \wedge b' \le b-1 \wedge i' \le i+(n-1)$,
$\trn_3 \equiv a' \le a \wedge b' \le b \wedge i' \le i-1, \quad \quad \quad
\trn_4 \equiv a' \le a-1 \wedge b' \le b+1 \wedge i' \le i-1$.\\
\CFA\ effectively merges loops at different control locations into a single loop creating one transition for every cyclic path of every loop (without unwinding inner loops).
This significantly simplifies the design of the later analysis phases.

\underline{\emph{3. Ranking Function Generation:}}
Our ranking function generation (Algorithm~\ref{alg:Termination} stated in Section~\ref{sec:termination}) finds an \emph{order} on the transitions resulting from \CFA\ such that there is a variable for every transition, which decreases on that transition and does not increase on the transitions that are lower in the order.
This results in the lexicographic ranking function $\lex = \langle a,a,b,i \rangle$ for the transitions $\trn_1,\trn_4,\trn_2,\trn_3$ in that order.
Our soundness theorem (Theorem~\ref{thm:termination-soundness}) guarantees that $\lex$ proves the termination of Figure~\ref{fig-ex1}.

\underline{\emph{4. Bound Analysis:}}
Our bound analysis (Algorithm~\ref{alg:Bound} stated in Section~\ref{sec:bound-computation}) computes a bound for every transition $\trn$ by adding for every other transition $\trnAlt$ how often $\trnAlt$ increases the variable of $\trn$ and by how much.
In this way, our bound analysis computes the bound $n$ for $\trn_2$, because $\trn_2$ can be incremented by $\trn_1$ and $\trn_4$, but this can only happen $n$ times, due to the initial value $n$ of $a$.
Further, our bound analysis computes the bound $n*(n-1)$ for $\trn_3$ from the fact that only $\trn_2$ can increase the counter $i$ by $n-1$ and that $\trn_2$ has the already computed transition-bound $n$.
Our soundness result (Theorem~\ref{thm:bounds-soundness}) guarantees that the bound $n$ obtained for $\trn_2$ is indeed a bound on how often the middle loop of Figure~\ref{fig-ex1} can be executed.

Our bound analysis solves the challenges (C1)-(C3):
\CFA\ allows us to analyze all loops at once (C1) creating one transition for every loop path (C2).
The abstract model of lossy VASS is precise enough to model counter increments, which is a key requirement for achieving amortized complexity analysis (C3).

\vspace*{-1mm}
\subsection{Amortized Complexity Analysis}
\label{subsec:amoritzed}

In his influential paper~\cite{amortizedComplexity} Tarjan introduces amortized complexity analysis using the example of a stack, which supports two operations \emph{push} (which puts an element on the stack) and \emph{popMany} (which removes several elements from the stack).
He assumes that the cost of {\em push} is 1 and the cost of {\em popMany} is the number of removed elements.
We use his example (see Figure~\ref{fig-amortized}) to discuss how our bound analysis achieves amortized analysis:
\begin{figure*}[t!]
\small
\begin{alltt}

  void main(int m) \{
     int i=m, n = 0;   //stack = emptyStack();
\(l\sb{1}\!:\)   while (i > 0) \{
        i--;
        if (?)  //push
           n++;        //stack.push(element);
        else    //popMany
\(l\sb{2}\!:\)         while (n > 0 && ?)
               n--;    //element = stack.pop();
\}   \}
\end{alltt}
\caption{Model of Tarjan's stack example \cite{amortizedComplexity} for amortized complexity analysis}
\label{fig-amortized}
\end{figure*}
Our analysis first abstracts the program to a VASS and then applies \CFA.
This results in the three transitions $\trn_1 \equiv i' = i - 1 \wedge n' = n + 1, \trn_2 \equiv i' = i - 1 \wedge n' = n, \trn_3 \equiv i' = i \wedge n' = n - 1$ (the first two transitions come from the outer loop, the last transition from the inner loop).
Algorithm~\ref{alg:Termination} then computes the lexicographic ranking function $\langle i,i,n \rangle$ for the transitions $\trn_1,\trn_2,\trn_3$ in that order.
Our bound analysis (Algorithm~\ref{alg:Bound}) then computes the joint bound $m$ for the transitions $\trn_1$ and $\trn_2$.
Our bound analysis further computes the bound $m$ for transition $\trn_3$ from the fact that only $\trn_1$ can increase the counter $n$ by 1 and that $\trn_1$ has the already computed bound $m$.
Adding these two bounds gives the amortized complexity bound $2m$ for Figure~\ref{fig-amortized}.
We highlight that our analysis has actually used the variable $n$ of transition $\trn_3$ as a \emph{potential function} (see~\cite{amortizedComplexity} for a definition)!
A lexicographic ranking function $\langle x_1,\ldots,x_n \rangle$ can be seen as a \emph{multidimensional potential function}.
Consider,  for example,  the ranking function $\langle a,a,b,i \rangle$ for the transitions $\trn_1,\trn_4,\trn_2,\trn_3$ of Figure~\ref{fig-ex1}.
The potential of $\trn_3$ can be increased by $\trn_2$ whose potential in turn can be increased by $\trn_1$ and $\trn_4$.

\section{Lossy VASSs and Basic Definitions}
\label{sec:program-model}

In this section we define lossy VASSs (introduced in~\cite{conf/stacs/BouajjaniM99}) and state definitions that we need later on.
We will often drop the `lossy' in front of `VASS' because we do not introduce non-lossy VASSs and there is no danger of confusion.
In this paper, we will use VASSs as \emph{minimal program model} for bound analysis of sequential programs without procedures.
We leave the extension to concurrent and recursive programs for future work.

\begin{definition}[Lossy Vector Addition System with States (VASS)]
\label{def:vass}
  We fix some finite set of \emph{variables} $\vars = \{x_1,\ldots,x_n\}$.
  A \emph{lossy vector addition system with states (VASS)} is a tuple $\prog = (\locs, \edges)$, where $\locs$ is a finite set of \emph{locations}, and $\edges \subseteq \locs \times \mathbb{Z}^n \times \locs$ is a finite set of \emph{transitions}.
  We write $\loc_1 \xrightarrow{\update} \loc_2$ to denote an edge $(\loc_1,\update,\loc_2)$ for some vector $\update \in \mathbb{Z}^n$.
  We often specify the vector $\update \in \mathbb{Z}^n$ by predicates $x_i' \le x_i + \update_i$ with $\update_i \in \mathbb{Z}$.

  A \emph{path} of $\prog$ is a sequence $\loc_0 \xrightarrow{\update_0} \loc_1 \xrightarrow{\update_1} \cdots$ with $\loc_i \xrightarrow{\update_i} \loc_{i+1} \in \edges$ for all $i$.
  A path is \emph{cyclic}, if it has the same start- and end-location.
  A path is \emph{simple}, if it does not visit a location twice except for start- and end-location.
  We write $\paath = \paath_1 \cdot \paath_2$ for the \emph{concatenation} of two paths $\paath_1$ and $\paath_2$, where the end-location of $\paath_1$ is the start-location of $\paath_2$.
  We say $\paath'$ is a \emph{subpath} of a path $\paath$, if there are paths $\paath_1$ and $\paath_2$ with $\paath = \paath_1 \cdot \paath' \cdot \paath_2$.

  The set of \emph{valuations} of $\vars$ is the set $V_\vars = \vars \rightarrow \mathbb{N}$ of mappings from $\vars$ to the natural numbers.
  A \emph{trace} of $\prog$ is a sequence $(\loc_0,\val_0) \xrightarrow{\update_0} (\loc_1,\val_1) \xrightarrow{\update_1} \cdots$ such that $\loc_0 \xrightarrow{\update_0} \loc_1 \xrightarrow{\update_1} \cdots$ is path of $\prog$, $\val_i \in V_\vars$ and $\val_{i+1}(x_j) \le \val_i(x_j) + \update_i$ for all $i$ and $1 \le j \le n$.
  $\prog$ is \emph{terminating}, if there is no infinite trace of $\prog$.
\end{definition}

Values of VASS variables are always non-negative.
We describe how to obtain VASSs from programs by abstraction in Section~\ref{sec:abstraction}.
The non-negativity of VASS values has two important consequences:
(1) Transitions in VASSs contain implicit guards:
for example a transition $x' \le x + \increment$ can only be taken if $x + \increment \ge 0$.
(2) VASS transitions can be used to model variable increments as well as variable resets:
we replace the assignment $x := k$, where $k \in \mathbb{Z}$, by the VASS transition $x' \le x + k$ during program abstraction (we point out that lossiness is essential for abstracting assignments).
This only increases the set of possible program traces and thus provides a conservative abstraction.

\paragraph{Parameterized VASSs.}
In our implementation we use a slight generalization of lossy VASSs.
We allow the increment $n$ in a transition predicate $x' \le x + n$ to be symbolic but constant; in particular, we require that $n$ does not belong to the set of variables $\vars$.
Our bound algorithm works equally well with symbolic increments under the condition that we know the sign of $n$.
We call these extended systems \emph{parameterized VASSs}.
See Figure~\ref{fig-ex1} for an example.

In the following we introduce some standard terminology that allows us to precisely speak about loops and related notions.
\begin{definition}[Reducible Graph, Loop Header, Natural Loop, Loop-nest Tree, e.g.~\cite{books/aw/AhoSU86}]
  Let $G = (V,E)$ be a directed graph with a unique entry point such that all nodes are reachable from the entry point.
  A node $a$ \emph{dominates} a node $b$, if every path from entry to $b$ includes $a$.
  An edge $\loc_1 \rightarrow \loc_2$ is a \emph{back edge}, if $\loc_2$ dominates $\loc_1$.
  $G$ is \emph{reducible}, if $G$ becomes acyclic after removing all back edges.
  A node is a \emph{loop header}, if it is the target of a back edge.
  The \emph{(natural) loop} of a loop header $h$ in a reducible graph is the maximal set of nodes $L$ such that for all $x \in L$ (1) $h$ dominates $x$ and (2) there is a back edge from some node $n$ to $h$ such that there is a path from $x$ to node $n$ that does not contain $h$.
\end{definition}

In the rest of this paper we restrict ourselves to VASSs and programs whose control flow graph is reducible.
This choice is justified by the fact that irreducible control flow is \emph{very rare} in practice (e.g. see the study in ~\cite{journals/spe/StanierW12}).
For analyzing irreducible programs we propose to use program transformations that make the program reducible; we do not elaborate this idea further due to lack of space.

Next, we define a special case of path, which corresponds to the notion of bound used in this paper (defined below).

\begin{definition}[Loop-path]
\label{def:loop-path}
  A \emph{loop-path} $\paath$ is a simple cyclic path, which starts and ends at some loop header $\loc$, and visits only locations inside the natural loop of $\loc$.
\end{definition}

{\em Example:}
$\loc_2 \xrightarrow{\trnAlt_2} \loc_3 \xrightarrow{\Id} \loc_2$ is a loop-path for the VASS in Figure~\ref{fig-ex1}.
However, $\loc_2 \xrightarrow{\Id} \loc_1 \xrightarrow{\trnAlt_1} \loc_2$ is not a loop-path because it does not stay inside the natural loop of $\loc_2$.
$\loc_2 \xrightarrow{\trnAlt_2} \loc_3 \xrightarrow{\Id} \loc_4 \xrightarrow{\trnAlt_4} \loc_3 \xrightarrow{\Id} \loc_2$ is not a loop-path, because it is not simple ($\loc_3$ is visited twice).

\begin{definition}[Instance of a loop-path]
\label{def:instance-loop-path}
Let $\paath = \loc_1 \xrightarrow{\update_1} \loc_2 \xrightarrow{\update_2} \cdots \loc_{n-1} \xrightarrow{\update_{n-1}}\loc_1$ be a loop-path.
A path $\paathL$ is an \emph{instance} of $\paath$ iff $\paathL$ is of the form $\loc_1 \xrightarrow{\update_1} \loc_2 * \loc_2 \xrightarrow{\update_2} \loc_3 * \loc_3 \cdots \loc_{n-1} * \loc_{n-1} \xrightarrow{\update_{n-1}} \loc_n = \loc_1$, where $\loc_i * \loc_i$ denotes any (possibly empty) path starting and ending at location $\loc_i$ which does not contain $\loc_1$.
A path $p$ \emph{contains} an instance $\paathL$ of $\paath$ iff $\paathL$ is a subpath of $p$.
Let be $\paathL$ be an instance of $\paath$ contained in $p$; a transition $t$ on $p$ \emph{belongs} to $\paathL$, if $t$ is on $\paathL$ and $t = \loc_i \xrightarrow{\update_i} \loc_{i+1}$ for some $i$.
\end{definition}

We note the following facts about instances:
Every transition in a path belongs to \emph{at most one} instance of a loop-path.
Every transition in a given cyclic path belongs to \emph{exactly one} instance of a loop-path.

{\em Example:}
There are four instances of loop-paths in the path $\paath = \loc_1 \xrightarrow{\trnAlt_1} \loc_2 \xrightarrow{\trnAlt_2} \loc_3 \xrightarrow{\Id} \loc_4 \xrightarrow{\trnAlt_3} \loc_3 \xrightarrow{\Id} \loc_2 \xrightarrow{\trnAlt_2} \loc_3 \xrightarrow{\Id} \loc_2 \xrightarrow{\Id} \loc_1$ of the VASS in Figure~\ref{fig-ex1}:
$\loc_1 \xrightarrow{\trnAlt_1} \loc_2 \xrightarrow{\Id} \loc_1$, $\loc_2 \xrightarrow{\trnAlt_2} \loc_3 \xrightarrow{\Id} \loc_2$ (twice) and $\loc_3 \xrightarrow{\Id} \loc_4 \xrightarrow{\trnAlt_3} \loc_3$.

\begin{definition}[Path-bound]
   A \emph{path-bound} for a loop-path $\paath$ is an expression $\bound$ over $\vars$ such that for every trace $(\loc_0, \val_0) \xrightarrow{\update_0} \cdots$ of $\prog$ the path $\loc_0 \xrightarrow{\update_0} \cdots$ contains at most $\bound(\val_0)$ instances of $\paath$.
\end{definition}

Path-bounds have various applications in bound and complexity analysis:
the \emph{computational complexity} of a program can be obtained by adding the bounds of the loop-paths of all loops;
a \emph{loop bound} can be obtained by adding the bounds of all loop-paths of a given loop;
the number of \emph{visits} to a \emph{single control location} $\loc$ can be obtained by adding the bounds of the loop-paths that include $\loc$ (our notion of a path-bound can be seen as a path-sensitive generalization of the notion of a ``reachability-bound''~\cite{conf/pldi/GulwaniZ10}); similarly one can obtain a bound on the number of \emph{visits} to a \emph{set of control locations}.
More generally, one can obtain resource bounds for a given \emph{cost model} by multiplying the bound on the number of visits to a control location with the \emph{cost} for visiting this location.

\begin{algorithm}[t!]
\SetKwFunction{cfa}{\CFA}
\SetKwInput{Procedure}{Procedure}
\Procedure{$\cfa(\prog)$}
\KwIn{a reducible VASS $\prog$}
\KwOut{a transition system $\atrns$}
\SetKwFunction{ComputeSummary}{Summary}
\SetKwFunction{ComputeSubstitutions}{ComputeSubstitutions}
\SetKw{Continue}{continue}

$\atrns$ := $\emptyset$\;  
\ForEach{loop header $\loc$ in $\prog$}{ 
\ForEach{loop-path $\paath = \loc \xrightarrow{\update_1} \loc_1 \cdots \loc_{n-1} \xrightarrow{\update_n} \loc$}{
  $\atrns$ := $\atrns \cup \{\update_1 + \cdots + \update_n\}$\;
  }}
\Return $\atrns$\;
\caption{{\tt \CFA} creates a {\em transition system} from a given VASS}
\label{alg:cfa}
\end{algorithm}

\section{Control Flow Abstraction}
\label{sec:transition-system}

\emph{Control flow abstraction} (\CFA), stated in Algorithm~\ref{alg:cfa}, is based on two main ideas:
(1) Given a program $\prog$, \CFA\ results into one transition for every loop-path $\paath$ for all loop ~ headers $\loc$ of $\prog$.
This enables a path-sensitive analysis, which ensures high precision during ranking function generation and bound analysis.
(2) The control structure is abstracted: effectively, all loops are merged into a single loop.
This allows to compute a common lexicographic ranking function for all loops later on.
\CFA\ maps VASSs to \emph{transition systems}.
Transition systems are not meant to be executed;
their sole purpose is to be used for ranking function generation and bound analysis.

\begin{definition}[Transition System]
  A \emph{transition system} is a set of vectors $\update \in \mathbb{Z}^n$.
  We often specify a transition $\update \in \mathbb{Z}^n$ by predicates $x_i' \le x_i + \update_i$, where $\update_i \in \mathbb{Z}$.
  We also write $\update \models x_i' \le x_i$ (resp. $\update \models x_i' < x_i$) for $\update_i \le 0$ (resp. $\update_i < 0$).
\end{definition}

\paragraph{Loop-path Contraction.}
Algorithm~\ref{alg:cfa} creates one transition for every loop-path $\paath = \loc \xrightarrow{\update_1} \loc_1 \cdots \loc_{n-1} \xrightarrow{\update_n} \loc$.
The transition $\update_1 + \cdots + \update_n$ represents the accumulated effect of all variable increments along the path.
The key idea of loop-path contraction is to \emph{ignore any inner loop} on $\paath$.
We will incorporate the effects of the inner loops only later on during the ranking function generation and bound analysis phase.

\paragraph{\CFA\ represents our choice of precision in the analysis:}
\CFA\ facilitates a high degree of disjunctiveness in the analysis, where we keep one disjunct for every loop-path.
By encapsulating the level of precision in a single analysis phase, we achieve a modular analysis (only during \CFA\ we need to deal with the control structure of the VASS).
This simplifies the design of the later termination and bound analysis and also allows us to easily adjust the analysis precision if the number of paths is prohibitively high (see the discussion on {\em path merging} in Appendix~\ref{sec:pathreduction}).

\section{Ranking Function Generation}
\label{sec:termination}

\begin{algorithm}[t!]
\SetKwInput{Procedure}{Procedure}
\Procedure{{\tt Ranking}$(\atrns)$}
\KwIn{a transition system $\atrns$}
\KwOut{a lexicographic ranking function $\lex$, which has one component for every transition $\trn \in \atrns$}
$\atrnsA := \atrns$\;
$\lex$ := ``lexicographic ranking function with no components''\;
\While{there is a transition $\trn \in \atrnsA$ and a variable $x$ such that $\trn \models x' < x$ and for all $\trn' \in \atrnsA$ we have $\trn' \models x' \le x$}{
  $\atrnsA$ := $\atrnsA \setminus \trn$\;
  $\lex$ := $\lex\mathit{.append}(x)$\;
}
\lIf{$\atrnsA = \emptyset$}{
  \Return $\lex$\;
}
\lElse{
  \Return ``Transitions $\atrnsA$ maybe non-terminating''\;
}
\caption{{\tt Ranking} computes a lexicographic ranking function}
\label{alg:Termination}
\end{algorithm}

In this section we introduce our algorithm for ranking function generation:
Algorithm~\ref{alg:Termination} reads in a transition system obtained from \CFA\ and returns a lexicographic ranking function that provides a witness for termination.
The key idea of the algorithm is to incrementally construct a lexicographic ranking function from local ranking functions.
We call a variable $x$ a \emph{local ranking function} for a transition $\trn$, if $\trn \models x' < x$.
A tuple $\lex = \left<y_1,y_2,\cdots,y_k\right> \in \vars^k$ is a \emph{lexicographic ranking function} for a transition system $\atrns$ iff for each $\trn \in \atrns$ there is a ranking function component $y_i$ that is a local ranking function for $\trn$ and $\trn \models y_j^\prime \le y_j$ for all $j < i$.
Algorithm~\ref{alg:Termination} maintains a candidate lexicographic ranking function $\lex$ and a set of transitions $\atrnsA$ for which no ranking function component has been added to $\lex$.
In each step the algorithm checks if there is a transition $\trn$ in $\atrnsA$ and a variable $x$ such that (1) $x$ is a local ranking function for $\trn$ and (2) no remaining transition increases the value of $x$, i.e.,
the condition $\forall \trn' \in \atrnsA. \trn' \models x' \le x$ is satisfied.
If (1) and (2) are satisfied, $\trn$ is removed from the set of remaining transitions $\atrnsA$ and $x$ is added as the component for $\trn$ in the lexicographic ranking function $\lex$.
Conditions (1) and (2) ensure that the transition $\trn$ cannot be taken infinitely often if only transitions from $\atrnsA$ are taken.
The algorithm stops, if no further transition can be removed.
If $\atrnsA$ is empty, the lexicographic ranking function $\lex$ is returned.
Otherwise it is reported that the remaining transitions $\atrnsA$ might lead to non-terminating executions.

Next we state the correctness of the combined application of Algorithm~\ref{alg:cfa} and Algorithm~\ref{alg:Termination}.
The proof can be found in Appendix~\ref{sec:proofs}.
\begin{theorem}
\label{thm:termination-soundness}
  If Algorithm~\ref{alg:Termination} returns a lexicographic ranking function $\lex$ for the transition system $\atrns$ obtained from Algorithm~\ref{alg:cfa} then VASS $\prog$ is terminating.
\end{theorem}

\paragraph{Reasons for Failure.}
There are two reasons why our ranking function generation algorithm may fail:
(1) There is a transition $\trn$ without a \emph{local ranking function}, i.e., there is no variable $x$ with
$\trn \models x' < x$.
Such a transition $\trn$ will never be removed from $\atrnsA$.
(2) There is a \emph{cyclic dependency} between local ranking functions, i.e., for every transition $\trn \in \atrnsA$ there is a local ranking function $x$ but the condition ``$\trn' \models x^\prime \le x$ for all $\trn' \in \atrnsA$'' is never satisfied.
We found cyclic dependencies to be very rare in practice (only 4 instances); we provide a discussion of the failures encountered in our experiments in Appendix~\ref{sec:limitations}.

\paragraph{Non-determinism.}
We note that in presence of transitions with more than one local ranking function, the result of Algorithm~\ref{alg:Termination} may depend on the choice for $x$.
However, it is straight-forward to extend Algorithm~\ref{alg:Termination} to generate all possible lexicographic ranking functions.

\section{Bound Computation}
\label{sec:bound-computation}

\begin{algorithm}[t!]
\SetKwInput{Procedure}{Procedure}
\SetKwInput{Global}{Global}
\Procedure{{\tt Bound}$(\trn)$}
\KwIn{a transition $\trn$}
\KwOut{a bound for transition $\trn$}
\Global{transition system $\atrns$, lexicographic ranking function $\lex$}
\SetKwFunction{InitialValue}{InitialValue}
\SetKwFunction{Bound}{Bound}
$x$ := ranking function component of $\trn$ in $\lex$\;
$\bound := \InitialValue(x)$\;
\ForEach{transition $\trn' \in \atrns$ with $\trn' \not\models x' \le x$ }{
  Let $k \in \mathbb{N}$ s.t. $x' \le x + k$ in $\trn'$\;
  $\bound := \bound + \Bound(\trn') \cdot k$\;
}
Let $k \in \mathbb{N}$ s.t. $x' \le x - k$ in $\trn$\;
\Return $\bound = \bound / k$\;
\caption{{\tt Bound} returns a bound for transition $\trn$}
\label{alg:Bound}
\end{algorithm}

In this section we introduce our bound algorithm:
Algorithm~\ref{alg:Bound} computes a bound $\bound$ for a transition $\trn$ of the transition system $\atrns$.
The main idea of Algorithm~\ref{alg:Bound} is to rely only on the components of the lexicographic ranking function $\lex$ for bound computation.
Let $x$ be the component of $\trn$ in $\lex$.
We recall that the termination algorithm has already established that $x$ is a local ranking function for $\trn$ and therefore we have $\trn \models x > x'$.
Thus $\trn$ can be executed at most $\InitialValue(x)$ often unless $x$ is increased by other transitions:
Algorithm~\ref{alg:Bound} initializes $\bound := \InitialValue(x)$ and then checks for every other transition $\trn'$ if it increases $x$, i.e., $\trn' \not\models x' \le x$.
For every such transition $\trn'$ Algorithm~\ref{alg:Bound} recursively computes a bound, multiplies this bound by the height of the increase $k$ and adds the result to $\bound$.
Finally, we divide $\bound$ by the decrease $k$ of $x$ on transition $\trn$.

\paragraph{Termination.}
Algorithm~\ref{alg:Bound} terminates because the recursive calls cannot create a cycle.
This is because Algorithm~\ref{alg:Bound} uses only the components of $\lex$ for establishing bounds and the existence of the lexicographic ranking function $\lex$ precludes cyclic dependencies.

\paragraph{Soundness.}
Our soundness result (Theorem~\ref{thm:bounds-soundness}, for a proof see Appendix~\ref{sec:proofs}) 
rests on the assumption that the CFG of $\prog$ is an SCC whose unique entry point is also its unique exit point.
We can always ensure this condition by a program transformation that encloses $\prog$ in a dummy while-loop
$\mathtt{while}(y > 0)\{ \prog\text{;} y\text{-{}-;}\}$,
where $y$ is a fresh variable with $\InitialValue(y) = 1$.
We point out that this program transformation enable us to compute path-bounds in terms of the program inputs for CFGs with multiple SCCs (e.g., a program with two successive loops).

\begin{theorem}
\label{thm:bounds-soundness}
  Let $\bound$ be a bound computed by Algorithm~\ref{alg:Bound} for a transition $\trn$ obtained from a loop-path $\paath$ during \CFA.
  Then $\bound$ is a path-bound for $\paath$.
\end{theorem}

\paragraph{Greedy Bound Computation.}
The bound computed by Algorithm~\ref{alg:Bound} depends on the lexicographic ranking function $l$.
Clearly, it is possible to run the algorithm for multiple lexicographic ranking functions and choose the minimum over the generated bounds.
However, we found the greedy approach to work well in practice and did not see a need for implementing the enumeration strategy.

\paragraph{Complexity of the Algorithm / Size of Bound Expressions.}
For ordinary VASS, the complexity of Algorithm~\ref{alg:Bound} is polynomial in the size of the input with a small exponent (depending on the exact definition of the complexity parameters).
Unfortunately, this statement does not hold for {\em parameterized} VASSs, for which bound expressions can be exponentially big:
We consider $n$ transitions $\trn_1, \ldots, \trn_n$ with the local ranking functions $x_1,\ldots,x_n$ and the lexicographic ranking function $\langle x_1,\ldots,x_n \rangle$.
We assume that transition $\trn_i$ increments $x_j$ by some constant $c_{ij}$ for $i < j$.
Then, Algorithm~\ref{alg:Bound} computes the bound stated in the following formula, which is exponentially big for symbolic coefficients $c_{ij}$:
\begin{displaymath}
\bound(\trn_n) = \sum_{k \in [0,n-1]} \prod_{i_1 < \cdots < i_k \newline \in [1,n-1]} \InitialValue(x_{i_1}) c_{i_1 i_2} \cdots c_{i_k n}
\end{displaymath}
However, in practical examples the variable dependencies are \emph{sparse}, i.e., most coefficients $c_{ij}$ are zero (confirmed by our experiments).
We highlight that Algorithm~\ref{alg:Bound} exploits this sparsity as it does not compute the bound using the explicit formula stated above but rather computes the bound for the current transition $\trn$ using only the bounds of the transitions that actually increase the counter of $\trn$ (i.e., $c_{ij} >0)$.
We note that in our experiments the computed bounds are small and the running time of Algorithm~\ref{alg:Bound} is basically linear in the number of transitions.
We conclude that in practice one should make use of the fine-grained precision offered by the possibly exponentially-sized bound expressions.

\paragraph{Preprocessing: Merging Transitions.}
Before the bound computation our analysis applies the following rule until no more transitions can be merged:
Let $\trn_1$ and $\trn_2$ be two transitions with the same local ranking function $x$ in $\lex$ such that $x' \le x + k \in \trn_1$ and $x' \le x + k \in \trn_2$ for some $k$ (i.e., both transitions decrement $x$ by the same amount).
We replace $\trn_1$ and $\trn_2$ by the transition $\trn = \{{y' \le y + \max\{k_1,k_2\}} \mid {{{y' \le y + k_1} \in \trn_1} \wedge {{y' \le y + k_2} \in \trn_2}}\}$.
It is not difficult to see that merging transitions is sound and always improves the bound computed by Algorithm~\ref{alg:Bound} (we do not give a formal proof here for lack of space).

\paragraph{Example:}
We have obtained the \emph{loop bound} of the middle loop in Figure~\ref{fig-ex1} from the path-bound $n$ of its single transition $\trn_2$ (see Section~\ref{sec:overview}).
We have obtained $2m$ as the \emph{amortized complexity} of Figure~\ref{fig-amortized} by adding the path-bounds of its transitions $\trn_1,\trn_2,\trn_3$ applying merging to $\trn_1$ and $\trn_2$ (see Section~\ref{subsec:amoritzed}).

\section{Program Abstraction}
\label{sec:abstraction}

In this section we describe how to abstract programs to VASSs.

\begin{definition}[Program] 
\label{def:program}
  Let $\states$ be a set of \emph{states}.
  The set of \emph{transition relations} $\trns = 2^{\states \times \states}$ is the set of relations over $\states$.
  A \emph{program} is a tuple $\prog = (\locs, \edges)$, where $\locs$ is a finite set of \emph{locations}, and $\edges \subseteq \locs \times \trns \times \locs$ is a finite set of \emph{transitions}.
  We write $\loc_1 \xrightarrow{\trn} \loc_2$ to denote a transition $(\loc_1,\trn,\loc_2)$.
  We assume the set of \emph{reachable states} $\Reach(\loc)$ is defined for every location $\loc \in \locs$ in the standard way.
  Let $e_1,e_2 \in \states \rightarrow \mathbb{Z}$ be integer-valued expressions over the states, and let $c \in \mathbb{Z}$ be some integer.
  We say $e_1 \ge 0$ is \emph{invariant for} $\loc$, if $e_1(\state) \ge 0$ holds for all $\state \in \Reach(\loc)$.
  We say $e_2' \le e_1 + c$  is \emph{invariant for} $\loc_1 \xrightarrow{\trn} \loc_2$, if $e_2(\state_2) \le e_1(\state_1) + c$ holds for all $(\state_1, \state_2) \in \rho$ with $\state_1 \in \Reach(\loc_1)$.
  We say $e_1$ is a \emph{norm}, if $e_1 \ge 0$ is invariant for every location $\loc$.
\end{definition}

\begin{definition}[Abstraction of a Program]
A VASS $V = (L, E^\prime)$ with variables $\vars$ is an \emph{abstraction} of a program $P = (L, E)$ iff
(1) every $x \in \vars$ is a norm and
(2) for each transition ${\loc_1 \xrightarrow{\rho} \loc_2} \in E$ there is a transition ${l_1 \xrightarrow{d} l_2} \in E^\prime$
such that every ${x^\prime \le x + c} \in d$ is invariant for $\loc_1 \xrightarrow{\rho} \loc_2$.
\end{definition}

The above definition suggests a three-step methodology for abstracting programs:
(1) Guess a set of norms $\candidates \subseteq \states \rightarrow \mathbb{Z}$.
(2) For every $x \in \candidates$ show that $x \ge 0$ is invariant at all locations $\loc$.
If this is not the case, discard the norm $x$.
(3) For every $x \in \candidates$ and every transition ${\loc_1 \xrightarrow{\rho} \loc_2}$ find a constant expression $c$ such that $x^\prime \le x + c$ is invariant for ${\loc_1 \xrightarrow{\rho} \loc_2}$.
Next, we describe how we implement this methodology.

\subsection{Abstracting Programs to VASSs: Our Implementation}
\label{subsec:program-abstraction}

\paragraph{Guessing Norms.}
The key idea of Algorithm~\ref{alg:Termination} is to find a local ranking function for every transition.
We recall that a transition is obtained from a loop-path during CA.
For this reason, our main heuristic is to consider expressions as norms that are local ranking functions for at least one loop-path of the program under analysis.
Our implementation iterates over all loop-paths $\paath = \loc \xrightarrow{\trn_1} \loc_1 \xrightarrow{\trn_2} \cdots \loc_{n-1} \xrightarrow{\trn_n} \loc$:
Let $\rel(\paath) = \trn_1 \circ \cdots \circ \trn_n$ be the transition relation obtained by contracting all transition relations along $\paath$.
We implement the computation of $\rel(\paath)$ by \emph{symbolic backward execution}, which returns a set of \emph{guards} $e \ge 0$ (we note that guards are normalized, e.g., $n \ge i$ is transformed into $n-i \ge 0$) and \emph{updates} $x' = e$, where $e$ is some expression over the program variables and $x'$ denotes the value of $x$ after executing $\paath$.
A \emph{local ranking function} is an expression $\rf$ such that (a) $\rf\ge 0$ is a guard of $\rel(\paath)$ and (b) $\delta_\rf = \rf - \rf' > 0$, where $\rf'$ denotes the expression $\rf$ where every variables $x$ is replaced by expression $e$ according to the update $x' = e$ of $\rel(\paath)$.
For every local ranking function $\rf$ our implementation adds the expression $\max\{\rf + \delta_\rf, 0 \}$ to the set of norms $\candidates$.
Clearly, all norms $x = \max\{\rf + \delta_\rf, 0 \} \in \candidates$ satisfy the invariant $x \ge 0$.

\paragraph{Abstracting Transitions.}
In our implementation we derive a transition predicate $x^\prime \le x + c$ for a given norm $x = \max\{e,0\} \in \candidates$ and transition ${\loc_1 \xrightarrow{\trn} \loc_2}$ as follows:
We obtain the expression $e'$ from $e$ by replacing variables with their updates according to $\trn$.
The expression $e'$ either constitutes an \emph{increment}, i.e., $e' = e + k_1$, or a \emph{reset}, i.e., $e' = k_2$, for some expression $k_i$.
For now, assume $k_i$ is constant.
We proceed by a case distinction:
If $e' = e + k_1$ and $e + k_1 \ge 0$ is invariant for ${\loc_1 \xrightarrow{\trn} \loc_2}$, then our implementation derives the transition predicate $x^\prime \le x + k_1$.
This derivation is sound, because of the invariant $e + k_1 \ge 0$.
(We motivate this derivation rule as follows: assume $\rf \ge 0$ and $\delta_\rf = \rf - \rf'$ hold on $\trn$, we have $e' = e + (- \delta_\rf) \ge 0$ for $e = \rf + \delta_\rf$.)
Otherwise, our implementation derives the transition predicate $x^\prime \le x + \max\{k_i,0\}$.
This derivation is sound because of properties of maxima.
If $k_i$ is not constant, we first search for an invariant $k_i \le u$ with $u$ constant, and then proceed as above (replacing $k_i$ with $u$).
We implement invariant analysis by symbolic backward execution (see Appendix~\ref{sec:proof-rules}).

\paragraph{Non-linear Local Ranking Functions.}
In our experiments we only found few loops that do not have a linear local ranking function.
However, these loops almost always involve the iterated division or multiplication of a loop counter by a constant such as in the transition relation $\trn \equiv x > 1 \wedge x' = x / 2$.
For such loops we can introduce the logarithm of $x$ as a norm, i.e., $y = \log x$, and then try to establish $y > 0$ from the condition $x > 1$ and derive the transition relation by $y' \le y - 1$ from the update $x' = x / 2$.

\paragraph{Data Structures.}
Previous approaches~\cite{conf/popl/GulwaniLS09,conf/popl/MagillTLT10} have described how to abstract programs with data structures to integer programs by making use of appropriate norms such as the length of a list or the number of elements in a tree.
In our implementation we follow these approaches using a light-weight abstraction based on optimistic aliasing assumptions.  
\begin{figure}[t!]{\scriptsize
\begin{tabular}{|l|l|l|l|l|l|l|l|l|l|l|l|}
\hline
		& { \textbf{Bounded}}		& $1$			& $\mathit{log} n$ 	& $n$ 		 & $n \mathit{log} n$ 	& $n^2$		 & $n^3$	 &	 $n^{> 3}$ & EXP & Time w/o Time-outs & \# Time-outs   \\\hline
Loopus 		& 383				& 131			& 0			& 151		 &	0		&	81	 &	16	 &	4	   &0	 & 437s & 5   \\\hline
KoAT		& 321				&121			&	0		&142		 &	0		&	54	 &	0	 &	3	   &0 	 & 682s & 282  \\\hline
PUBS		& 279				&116			&	5		&129		 &5			&15		 &4		 &	0	   &6	 & 1000s & 58 \\\hline
Rank  		& 84				&56			&	0		&19		 &0			&8		 &1		 &	0	   &0	 & 173s &  6   \\\hline
\end{tabular}}
\caption{Analysis results for the benchmark from \cite{conf/tacas/BrockschmidtEFFG14}.}
\vspace*{-4mm}
\label{AachenResults}
\end{figure}

\section{Experiments}
\label{sec:experiments}

We implemented the discussed approach as an intraprocedural analysis (we use function inlining) based on the LLVM~\cite{LLVM} compiler framework.
Our tool \loopus computes loop bounds (depending on a command-line parameter, see~\cite{loopuswebsite}) either in terms of
(1) the inputs to the SCC to which the loop belongs, or
(2) the function inputs (this is implemented by enclosing the function body in a dummy loop as described in Section~\ref{sec:bound-computation}).
At the same time \loopus also computes the asymptotic complexity of the considered SCC.
\loopus works only for reducible control flow graphs as discussed in Section~\ref{sec:program-model}.
\loopus models integers as mathematical integers (not bit-vectors), which is the standard approach in the bound analysis literature.
We use the Z3 SMT solver~\cite{DeMoura:2008} for removing unsatisfiable paths during the analysis.
Given a loop condition of form $a \ne 0$ \loopus heuristically decides to either assume $a > 0$ or $a < 0$ as loop-invariant;
this assumption is reported to the user.
Similarly, \loopus assumes $x > 0$ when an update of a loop counter of the form $x = x * 2$ or $x = x / 2$ is detected.
The task of validating these assumptions is orthogonal to our approach and can be performed by standard tools for invariant generation.
\loopus and more details on our experimental evaluation are available at~\cite{loopuswebsite}.
 
\subsection{Comparison to Tools from the Literature}

We compare \loopus against the tools \koat\cite{conf/tacas/BrockschmidtEFFG14}, \pubs\cite{journals/tcs/AlbertAGPZ12,AlbertGM12} and \rank\cite{conf/sas/AliasDFG10}.
For the comparison we use the benchmark~\cite{aachenbench},
which consists of small example programs from the bound analysis literature and the benchmark suite which was used to evaluate \terminator~\cite{conf/cav/BrockschmidtCF13}.
Since \loopus expects C code but \koat expects a transition system as input, we needed to obtain C programs for comparison.
For the examples from the literature we used a script and hand-translated the programs to C when the script was not sufficient.
In particular, we translated tail-recursive programs to while-programs as our implementation does not support recursion (tail-recursion would be an easy extension).
We excluded the RAML programs from the benchmark, because we do not see a meaningful way to compare C translations to functional programs.
For \terminator we used the goto-programs from~\cite{conf/cav/BrockschmidtCF13}, from which the transition systems in~\cite{aachenbench} were created (the original benchmark from which the goto-programs were created is not available).
Figure~\ref{AachenResults} states the results for the different tools (the results for {\tt KoAT}, {\tt PUBS} and {\tt Rank} were taken from \cite{conf/tacas/BrockschmidtEFFG14}).
Columns 2 to 9 state the number of programs that were found to have the given complexity by the respective tool.
The table shows that \loopus can compute bounds for more loops than the other tools.
Moreover one can see significant differences in analysis time, which are due to time-outs (the table shows the analysis time without time-outs and the number of time-outs separately; the time-out is set to 60s for all tools).
The detailed comparison available at~\cite{loopuswebsite} shows that there are 84 loops for which \loopus computes an asymptotically more precise bound than any of the 3 other tools, compared to 66 loops for which one of the 3 other tools computed an asymptotically more precise bound than \loopus.

\subsection{Evaluation on Real-World Code}

\begin{figure}[t!]{\scriptsize
\begin{tabular}{|l|l|l|l|l|l|l|l}
\hline
		& { \textbf{Analyzed}}				& { Outer Dep.}				& {Inner Dep.} 				 & {Paths $>$ 1} 			 & {Non-Trivial} \\\hline
Loops 		& \ 4210\hfill	\ 	   			&\ 255\hfill	\ 			&\ 305	\ 				 &	\ 1276\hfill	\ 		& \ 1475\hfill		      \\
Bounded		& \ 3205\hfill	[3060]\ 	   		&\ 120\hfill	[112]\ 			&\ 148	 [129]\ 			 &	\ 744\hfill	 [695]\ 		& \ 831\hfill	[812]	      \\
  / Overall	& \ 76\%\hfill 	[73\%]\ 	   		&\ 47\%\hfill	[44\%]\ 		&\ 49\%	 [42\%]\ 			 &	\ 58\%\hfill	 [54\%]\ 	& \ 56\%\hfill	[55\%]	      \\\hline
SCCs  		& \ 2833	 	   			& \ 181					&\ 193 					 &	 \ 902				& \ 937		      \\
Bounded		& \ 2289		   			& \ 70					&\ 95					 &	 \ 542				& \ 564		      \\
  / Overall	& \ 81\%		   			& \ 39\%				&\ 50\%					 &	 \ 60\%				& \ 60\%		      \\\hline
\end{tabular}}
\caption {Loop and SCC Statistic of our current implementation for the cBench Benchmark, the results obtained with the implementation of \cite{conf/sas/ZulegerGSV11} are given in square brackets}
\vspace*{-4mm}
\label{cBenchResults}
\end{figure}

We evaluated \loopus on the program and compiler optimization benchmark {\em Collective Benchmark}~\cite{cbench}(cBench), which contains a total of 1027 different C files (after removing code duplicates) with 211.892 lines of code.

\paragraph{Data Structures.}
For expressing bounds of loops iterating over arrays or recursive data structures \loopus introduces shadow variables representing appropriate norms such as the length of a list or the size of an array.
\loopus makes the following optimistic assumptions which are reported to the user:
Pointers do not alias, a recursive data structure is acyclic if a loop iterates over it,
a loop iterating over an array of characters is assumed to be terminating if an inequality check on the string termination character '\textbackslash 0' is found.\footnote{This assumption is necessary since the type system of C does not distinguish between an array of characters and a string}
We made these assumptions in order to find interesting examples, a manual check on a sample of around 100 loops in the benchmark found the assumptions to be valid with respect to termination.
The task of validating an assumption is orthogonal to our approach and can be performed by standard tools for shape analysis.

\paragraph{Results.}
In Figure~\ref{cBenchResults}, we give our results on different loop classes.
We recall that our bound analysis is based on an {\em explicitly} computed termination proof.
We do not list the results of our termination analysis separately, because a bound was computed for 98\% of all loops for which termination was proven.
Details on the reasons for failure of our termination analysis and bound analysis, which occurred during the experiments, are given in Appendix~\ref{sec:limitations}.
In column {\em Analyzed} we state the results over all loops in the benchmark.
We summarize the results over all loop categories except {\em Analyzed} in the column {\em Non-Trivial}.

\paragraph{Challenging Loop Classes.}
We defined syntactic categories for loops that deviate from standard for-loops '\texttt{for(i = 0; i < n; i++)}'.
These categories pose a challenge to bound analysis tools.
The loop-class `{\em Outer Dependent}' captures all outer loops whose termination behavior is affected by the executions of an inner loop.
(E.g., in Figure~\ref{fig-ex1} termination of loop $l_1$ depends on loop $l_3$, while in Figure~\ref{fig-amortized} termination of loop $l_1$ does {\em not} depend on loop $l_2$.)
We define an inner loop to be in the set of `{\em Inner Dependent}' loops if it has a loop counter that is {\em not} reset before entering the loop.
(E.g., in Figure~\ref{fig-ex1} loop counter $i$ of loop $l_3$ is always reset to $n-1$ before entering the loop, while loop counter $b$ of loop $l_2$ is never reset.)
The loop-class '{\em Paths $>$ 1}' contains all loops which have more than 1 path left after program slicing (see Appendix~\ref{sec:pathreduction}).
The categories for the SCCs are the same as for the loops: we define an SCC to be in a certain category if it contains at least one loop which is in that category.
Success ratios of around 50\% in the difficult categories demonstrate that our method is able to handle non-trivial termination and complexity behavior of real world programs.

\paragraph{Amortization.}
For 107 loops out of the 305 loops in the class `{\em Inner Dependent}', the bound that our tool computed was amortized in the sense that it is asymptotically smaller than one would expect from the loop-nesting depth of the loop.
In 12 cases the amortization was caused by incrementing a counter of the inner loop in the outer loop as in Figures~\ref{fig-ex1} and \ref{fig-amortized}.
The 12 loops are available at \cite{loopuswebsite}.
\emph{For these loops a precise bound cannot be computed by any other tool (as discussed in the beginning of Section~\ref{sec:overview}).}

\paragraph{Performance.}
The results were obtained on a Linux machine with a 3.2 Ghz dual core processor and 8 GB Ram.
92 loops of the 4302 loops in our benchmark are located in 44 SCCs with an irreducible control flow.
We thus analyzed 4210 loops.
The total runtime of our tool on the benchmark (more than 200.000 LoC) did not exceed 20 minutes.
The time-out limit of maximal 420 seconds computation time per SCC was not reached.
There were only 27 out of 2833 SCCs (174 out of 4210 loops) on which the analysis spent more than 10 seconds.

\paragraph{Experimental Comparison.}
We compared our tool against our previous work~\cite{conf/sas/ZulegerGSV11}, which to the best of our knowledge represents the only other experimental evaluation of bound analysis on a large publicly-available benchmark of C programs.
For the purpose of a realistic comparison, we ran the tool of \cite{conf/sas/ZulegerGSV11} on the same machine with an equal time out limit of 420 second.
The results are given in square brackets in Figure~\ref{cBenchResults}.
Note the significant increase in the number of loops bounded in each of the challenging categories.
The execution of the tool \cite{conf/sas/ZulegerGSV11} took an order of magnitude longer (nearly 13 hours) and we got 78 time-outs.
The main reason for the drastic performance increase is our new reasoning on inner loops:
The approach of \cite{conf/sas/ZulegerGSV11} handles inner loops by inserting the transitive hull of an inner loop on a given path of the outer loop.
This can blow up the number of paths exponentially.
We avoid this exponential blow-up thanks to \CFA:
\CFA\ allows us to analyze inner and outer loops at the same time and thus eliminates the need for transitive hull computation.
The only complexity source of our approach is its path-sensitivity.
Thanks to the simplicity and modularity of our method, we can tackle the path explosion problem by simple path reduction techniques (see Appendix~\ref{sec:pathreduction}).

\paragraph{Acknowledgements.} 
We thank Fabian Souczek and Thomas Pani for help with the experiments.

\bibliographystyle{plain}
\bibliography{main}

\newpage

\appendix

\section{Proofs}
\label{sec:proofs}

\begin{proof}[of Theorem~\ref{thm:termination-soundness}]
  Assume Algorithm~\ref{alg:Termination} returns a lexicographic ranking function $\lex$.
  For the sake of contradiction we assume that there is an infinite trace $(\loc_0,\val_0) \xrightarrow{\update_0} (\loc_1,\val_1) \xrightarrow{\update_1} \cdots$ of $\prog$.
  Thus there is at least one loop-path with infinitely many instances in the path $\paath = \loc_0 \xrightarrow{\update_0} \loc_1 \xrightarrow{\update_1} \cdots$.
  By definition of \CFA\ $\atrns$ contains one transition per loop-path of $\prog$.
  Given a transition $\trn$, we denote by $\phi(\trn)$ the loop-path from which $\trn$ has been obtained.
  Now, we fix a transition $\trn$ that is the minimal transition in $\lex$ such that $\phi(\trn)$ has infinitely many instances in $\paath$.
  Let $k$ be an index in $\paath$ such that all loop-path that have only finitely many instances appear before $k$.
  Let $x$ be the ranking function component of $\trn$ in $\lex$.
  We denote by $x' \le x + c_i$ the transition predicate of $x$ in $\update_i$.
  Let $k \le k_1 < k_2 < \cdots$ be the start indices of the infinitely many instances of $\trn$ in $\paath$.
  We consider the accumulated effect of $x$ between locations $\loc_{k_i}$ and $\loc_{k_{i+1}}$.
  We have $x_{i+1} \le x_i + c_{k_i} + c_{k_i + 1} + \cdots + c_{k_{i + 1} - 1}$ (*).
  We note that the path $\loc_{k_i} \xrightarrow{\update_{k_i}} \cdots \xrightarrow{\update_{k_{i + 1} - 1}} \loc_{k_{i + 1}}$ is cyclic and therefore can be completely decomposed into instances of loop-paths.
  We now reorder the sum (*) in order to aggregate the summands which belong to instances of the same loop-path (this is justified by the fact that plus is commutative): $x_{i+1} \le x_i + n_{\trn'} \cdot c_{\trn'} + n_{\trn''} \cdot c_{\trn''} + \cdots$ (**), where $c_{\trn*}$ denotes the total of the summands of the loop-path $\phi(\trn*)$ and $n_{\trn*}$ denotes the number of instances of the loop-path $\phi(\trn*)$.
  Because $\trn$ has been chosen minimal we have that no transition $\trn*$ appears before $\trn$ in $\lex$, and we get $\trn* \models x' \le x$.
  Therefore, $c_{\trn*} \le 0$ for all $\trn*$.
  We know that $\trn$ appears in (**) and that $c_\trn < 0$ because $x$ is the local ranking function of $\trn$ in $\lex$.
  Thus we get $x_{i+1} < x_i$ from (**).
  Because that holds for all $i$ we get an infinitely decreasing chain.
  This is impossible because all $x_i$ are non-negative by assumption.
\end{proof}

\begin{proof}[of Theorem~\ref{thm:bounds-soundness}]
By definition of \CFA\ $\atrns$ contains one transition per loop-path of $\prog$.
Given a transition $\trn$, we denote by $\phi(\trn)$ the loop-path from which $\trn$ has been obtained.
Let $\trn$ be a transition of the VASS $\atrns$.
Let $x$ be the ranking function component of $\trn$ in $\lex$.
Let $(\loc_0,\val_0) \xrightarrow{\update_0} (\loc_1,\val_1) \xrightarrow{\update_1} \cdots$ be a trace of $\prog$ that starts and ends at the unique entry point of $\prog$.
We denote by $x' \le x + c_i$ the update of $x$ in $\update_i$.
By definition of a VASS, variables take values only in the non-negative numbers.
In particular, the final value of $x$ is non-negative, i.e., $\InitialValue(x) + c_0 + c_1 \cdots \ge 0$ (*).
Because $\loc_0 \xrightarrow{\update_0} \loc_1 \xrightarrow{\update_1} \cdots$ is cyclic the sequence (*) can be completely decomposed into instances of loop-paths.
Because plus is commutative we can reorder the sequence (*) and aggregate the summands that belong to instances of the same loop-path:
$\InitialValue(x) + n_{\trn} \cdot c_\trn + n_{\trn'} \cdot c_{\trn'} + n_{\trn''} \cdot c_{\trn''} + \cdots \ge 0$ (**), where $c_{\trn*}$ denotes the total of the summands of the loop path $\phi(\trn*)$ and $n_{\trn*}$ denotes the number of instances of the loop path $\phi(\trn*)$.
Algorithm~\ref{alg:Termination} has established $c_\trn < 0$ and $c_{\trn*} \le 0$ for all transitions $\trn*$ that appear after $\trn$ in $\lex$.

We now prove the claim by induction on the position of
$\trn$ in the lexicographic ranking function $\lex$.
Base case: $\trn$ is first and we get $\InitialValue(x) + n_\trn \cdot c_\trn \ge 0$ from (**) and thus $n_\trn \le \InitialValue(x)$.
Step case: by induction assumption we have $n_{\trn*} \le \Bound(\trn*)$ for all transitions $\trn*$ that appear before $\trn$ in $\lex$.
We get $\InitialValue(x) + n_\trn \cdot c_\trn + \Bound(\trn') \cdot c_{\trn'} + \Bound(\trn'') \cdot c_{\trn''} + \cdots \ge 0$ from (**) and thus $n_\trn \le  (\InitialValue(x) + \Bound(\trn') \cdot c_{\trn'} + \Bound(\trn'') \cdot c_{\trn''} + \cdots ) / c_\trn$.
\end{proof}

\section{Invariant Generation by Symbolic Backward Execution}
\label{sec:proof-rules}

We implement invariant generation based on symbolic backward execution as described in~\cite{backward_symb_exec}.
The cited approach allows to derive invariants in a compositional and demand-driven manner.
In order to make the paper more self-contained we sketch here the techniques from~\cite{backward_symb_exec} in terms of four proof rules.
The proof rules describe how to derive an upper bound invariant $e \le u$ for a given expression $e$ at a given program location $\loc$; symmetric rules can be given for lower bounds $u \le e$. 

\textbf{(1)} If $e$ is a composed linear expression, e.g.,  $e = 2x - 3y$, we recursively compute an upper bound $u_1$ for $x$ and a lower bound $u_2$ for $y$ and then compose these bounds to the bound $2u_1 - 3u_2$ for $e$.

\textbf{(2)} If there is a condition $e \le u$ at some location $\loc'$ dominating location $\loc$ and $e$ is not modified between $\loc'$ and $\loc$, we obtain the invariant $e \le u$, if $u$ is constant.
Otherwise, we try to recursively obtain an upper bound for $u$.

\textbf{(3)} If $e$ is never increased throughout the program, we obtain the invariant $e \le \InitialValue(e)$.

\textbf{(4)} If $e$ is a variable, we iterate over all reaching definitions $e = e'$;
if a reaching definition is an increment of $e$ inside some loop, i.e., $e' = e + c$, we apply our bound analysis algorithm to establish an upper bound $b$ on how often $e = e + c$ can be executed; otherwise, we recursively obtain an upper bound for $e'$; then we return the maximum over these upper bounds plus $b*c$ for every increment.
We illustrate this rule on Example~\ref{ex:invariant1}:
Assume we want to compute an upper bound for $j$ at $\loc_2$.
The only reaching definition to $\loc_2$ comes from the assignment $j = a$ on the outer loop.
$a$ at this assignment has two reaching definitions: $a = m$ and the increment $a = a + 4$.
Proof rule 5 now applies our bound analysis algorithm for establishing the bound $n$ on how often $a = a + 4$ can be executed.
This results in the invariant $a \le m + n * 4$.


\begin{figure}
\begin{example}
\label{ex:invariant1}
\begin{alltt}

void main(uint n, uint m) \{
    uint a = m; uint i = 0;
\(l\sb{1}\!:\)  while(i < n) \{
       j = a;
\(l\sb{2}\!:\)     while(j > 0)
          j--;
       a = a + 4; i++;
\}   \}
\end{alltt}
\end{example}
%
%
%
%
%
%
\vspace*{-4mm}
\label{fig-ex3}
\end{figure}

\section{Path Reduction}
\label{sec:pathreduction}
\begin{figure}\scriptsize
 \begin{tabular}{| c | c|c|c|c|c|c|c|c|c|c}
\hline
{\# paths}		& {1}	& {2} 	& {3 - 9}	& {10 - 99}	& {100 - 299} 	& {300 - 1999} 	& {2000 - 4999} 		 & {$\ge$ 5000}  \\\hline
{unsliced}	& 1174		& 578 		& 616 		& 310 			& 66 					 & 44 					 & 11 			 	 		 & 34	\\\hline
{sliced}  	& 1623		& 512		& 424		& 183 			& 37 					 & 30					 & 8 			 	 		 & 16	\\\hline
{merged}  	& 1766		& 429 		& 415 		& 186 			& 24 					 & 10 					 & 1 			 	 		 & 2	\\\hline
{refined} 	& 1766		& 429		& 414 		& 187			& 24 					 & 10 					 & 1 			 	 		 & 2	\\\hline
 \end{tabular}
\caption{Number of paths with respective number of SCCs }
\vspace*{-4mm}
\label{fig:exppaths}
\end{figure}
First we apply {\em program slicing} with regard to the loop exit conditions, i.e. we delete all program behavior that cannot affect loop termination.
In the next step we exclude path doubles through syntactic comparison.
On loops with more than 250 paths left, we apply what we call {\em path merging}:
For each path $p$ the conjunction $c_p$ over all its predicates is built. Paths which assign syntactically identical expressions to the
loop counters are grouped.
The paths in the same group $G$ are substituted by a new path with the single predicate $\bigvee_{p \in G} c_p$ (we simplify this predicate by standard techniques from propositional logic).
The merged path overapproximates all paths in $G$.
Though some path sensitivity is lost by this technique, we can still bound 81 (31\%) of the 259 loops to which path merging is applied.
After the second path reduction step we apply standard control flow refinement techniques such as loop unrolling.
Figure~\ref{fig:exppaths} states the number of SCCs with the respective number of paths in the original program (first row), in the sliced program (second row),
after deleting path duplicates and applying path merging (third row)
and after applying control flow refinement (last row). We state the paths per SCC because in our implementation all loops in one SCC are processed at once.
Our statistics (Figure~\ref{fig:exppaths}) demonstrate that slicing and merging significantly reduce the number of paths while control flow refinement does not lead to any
problematic increase in the path count.

\section{Limitations}
\label{sec:limitations}
For a total of 967 loops our implementation failed to prove termination.
For 38 loops our tool proved termination but was not able to infer a bound.
The main reason for not being able to compute a bound is that we cannot infer an upper bound invariant for a variable reset, e.g., a variable $x$ is reset to some non-constant expression $e$ and the techniques discussed in Section~\ref{sec:proof-rules} are not sufficient to establish a constant upper bound $u$ on $e$.
This typically happened due to a missing heap invariant (e.g. all array elements are smaller than $n$).
We distinguish the two reasons for failure of our termination analysis discussed in Section~\ref{sec:termination}:

\paragraph{No local ranking function.}
For 903 loops our analysis failed because there was a loop-path with no local ranking function.
We analyzed on a random sample of 50 loops out of the 903 loops the reasons for failure.
The first class of failures is due to insufficient modeling features of our implementation: bitwise operations (not modeled, 5 cases), function inlining (applied very restrictive, 4 cases), unsigned integers (modeled as integers, 2 cases), external functions (not modeled, 4 cases).
In 5 cases termination is conditional and cannot be proven (e.g., a character stream must contain the line break character).
In 7 cases function pointers need to be resolved.
In 27 cases our invariant analysis is insufficient (7 array invariants, 20 arithmetic invariants).

\paragraph{Cyclic dependencies.}
For a total of 64 loops, our analysis found a local ranking function for every loop-path but was not able to compute a lexicographic ranking function, because of {\em cyclic dependencies}.
However, a manual inspection of the 64 loops revealed that only 4 cases are indeed instances of a cyclic dependency failure.
For the other 60 loops the real reason for failure is that our tool was not able to compute the right local ranking function for certain loop-paths (for the same reasons discussed above); the reported cyclic dependency was caused by variables that were actually not relevant for the termination of the loop.
In the 4 remaining cases control flow refinement by {\em contextualization} \cite{conf/sas/ZulegerGSV11} would resolve the cyclic dependency.

\end{document}